\documentclass[conference]{IEEEtran}
\IEEEoverridecommandlockouts
\usepackage{cite}
\usepackage{amsmath,amssymb,amsfonts}
\usepackage{algorithmic}
\usepackage{graphicx}
\usepackage{textcomp}
\usepackage{xcolor}
\usepackage{comment}
\usepackage{hyperref}
\usepackage{xurl}   % va caricato dopo hyperref
\def\BibTeX{{\rm B\kern-.05em{\sc i\kern-.025em b}\kern-.08em
    T\kern-.1667em\lower.7ex\hbox{E}\kern-.125emX}}
    
\definecolor{darkgreen}{RGB}{0, 102, 0}

\begin{document}

\title{A Wireless Network Architecture for Monitoring of Hospitalized Patients %A network architecture for early warning and real-time monitoring of hospitalized patients in pre-intensive care
\thanks{
\copyright~2024 IEEE. Personal use of this material is permitted.
Permission from IEEE must be obtained for all other uses, in any
current or future media, including reprinting/republishing this
material for advertising or promotional purposes, creating new
collective works, for resale or redistribution to servers or lists,
or reuse of any copyrighted component of this work in other works.
This is the accepted version of a paper published in \emph{25th IEEE International Conference on Emerging Technologies and Factory Automation (ETFA), Vienna, Austria, 2020}. The final published
version is available at:
\href{https://doi.org/10.1109/ETFA46521.2020.9211950}{https://doi.org/10.1109/ETFA46521.2020.9211950}

This research is funded by the INTERREG V-A Italy-Malta Programme 2014 - 2020, through the “MEDIWARN – Virtual Biosensor for Medical Warning Precursors” European Project (CUP E69F18000000005).}
}

\author{\IEEEauthorblockN{{Luca Leonardi, Lucia Lo Bello, Gaetano Patti}}
\IEEEauthorblockA{\textit{Department of Electrical, Electronics and Computer Engineering} \\
\textit{University of Catania}, Catania, Italy \\
{\{luca.leonardi, lobello, gaetano.patti\}}@unict.it
}}

\maketitle

\begin{abstract}
Monitoring of vital parameters plays a key role in better understanding the clinical condition of hospitalized patients.
In this perspective, the virtual biosensor for MEDIcal WARNing precursor (MEDIWARN) aims to help the medical team to take a prompt action in response to the evolution of  patients' health condition. 
This paper presents the MEDIWARN network architecture and discusses the design challenges that such a system poses. 
Simulative assessments provide a preliminary evaluation of the system in realistic scenarios.
\end{abstract}

\begin{IEEEkeywords}
Monitoring systems, Network architecture, Wireless networks.% mesh networks, ad-hoc networks, IEEE 802.11, ...
\end{IEEEkeywords}

\section{Introduction and Related Work}
\label{sec:introduction}
Recent IoT and AI-based technologies help caregivers in providing continuous support to their patients, both at home and when they are in the hospital. In the first case, vision-based applications allow to detect critical and dangerous situations~\cite{5514566} while preserving the patients' privacy. In the second case, when more severe conditions require the hospitalization of the patients, continuous monitoring of their vital signs (e.g., blood pressure, body temperature, etc.) allows to store the clinical evolution of the patients and process the relevant data, thus ensuring early action in case of a progressive clinical deterioration~\cite{2_MEDIWARN,8_MEDIWARN}.
In this context, multiple research projects aim to propose new technologically advanced solutions to support clinical medicine in the study and analysis of vital parameters of hospitalized patients~\cite{1_MEDIWARN}. Moreover, several works in the literature, such as~\cite{IoT_7955637,IoT_7917920,LoRaBioMedical,MEDIWARNelectronicsLoRaWAN,patientMonitoringRaspberry,AMIN2018483} propose monitoring systems for healthcare based on Internet-of-Thing technologies.  
Nowadays, the Early Warning Score (EWS)~\cite{bonnici2015evaluation} is the commonly used indication of the patient health status. The EWS uses multiple sensors to collect physiological parameters and produce a score that allows the doctors to quickly define the urgency level of a patient. However, the EWS is not able to foresee alert conditions and therefore predict any upcoming critical condition for the patient. In this direction, the virtual biosensor for MEDIcal WARNing precursor (MEDIWARN)~\cite{MEDIWARN_8805005} realizes a system for continuous monitoring of the vital parameters (i.e., body temperature, blood pressure, respiratory rate and oxygen saturation) of hospitalized patients that includes a smart warning system able to predict a possible medical alert and warn about the deterioration of the patient’s condition.

The MEDIWARN system requires a network infrastructure that allows to collect sensor data in a central station that processes data in real time using mathematical models and advanced fuzzy logic. 
In generic hospital units, a communication network infrastructure may not be available. For this reason, MEDIWARN relies on a wireless network architecture that does not require any wired infrastructure.

This work presents the MEDIWARN network architecture, discussing the relevant  design challenges and the results of simulative assessments in realistic scenarios. 

The paper is organized as follows. Section~\ref{sec:architecture} presents the MEDIWARN network architecture and the relevant design challenges. 
Section~\ref{sec:simulativeAssessment} addresses a simulative assessment of the proposed architecture.
Finally, Section~\ref{sec:conclusions} gives conclusions and hints for future works.

\section{The MEDIWARN System Architecture}
\label{sec:architecture}

\begin{figure}[tb]%[htbp]
\centerline{\includegraphics[scale= 0.46]
{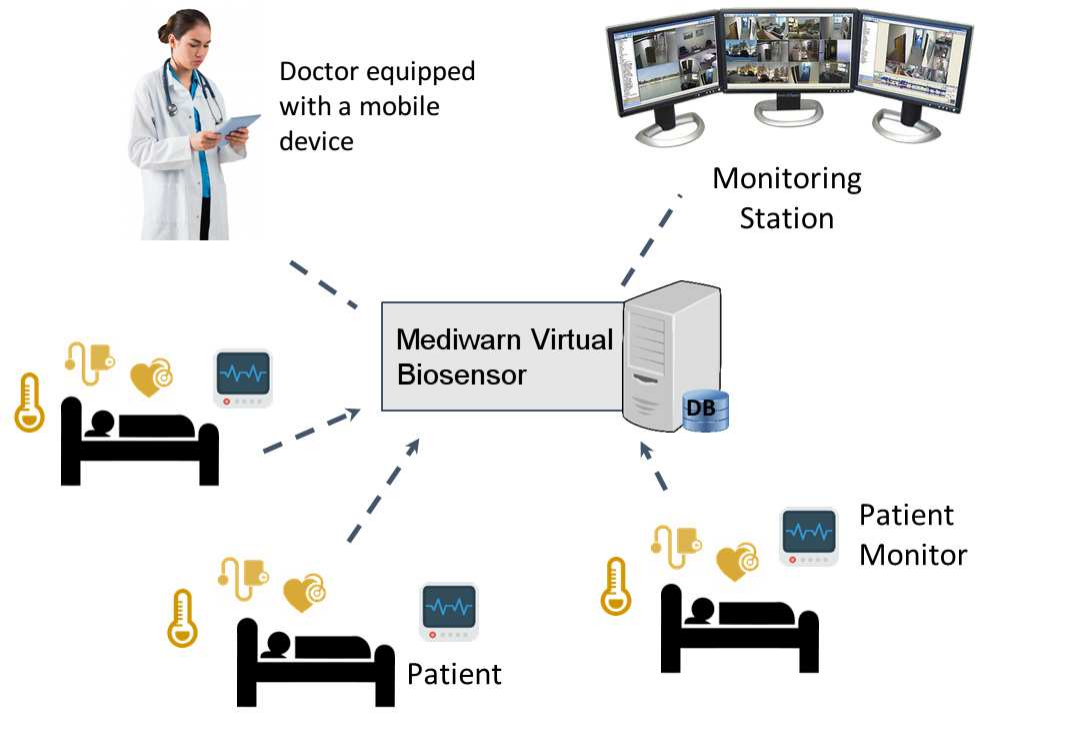}
}
\caption{MEDIWARN network architecture.}
\label{fig:MEDIWARNarchitecture}
\end{figure}

The MEDIWARN system architecture is shown in Fig.~\ref{fig:MEDIWARNarchitecture}. 
Each monitor consists of a set of sensors for monitoring the patient vital parameters. These parameters are shown on a screen that is close to the patient and they are also transmitted to the MEDIWARN Virtual Biosensor, i.e., the server that stores and processes the acquired vital parameters. Such a server maintains a database with all the vital parameters sampled from the patients and runs the predictive algorithm that will alert the medical staff if the patient status is about to get worse. The output of the %MEDIWARN Virtual Biosensor
system is shown on both:
\begin{itemize}
    \item The monitoring station, which gives a detailed view of each patient's condition.
    \item The handheld devices (e.g., tablets) of the medical staff that provide a summarized view of the patient status.
\end{itemize}

%{\color{blue}
Both the monitoring station and the handheld devices periodically poll the database server to show the current patients' status and the outcome of the smart warning system implemented on the MEDIWARN Virtual Biosensor.
%}
%Both the monitoring station and the handheld devices poll the database server at regular and configurable intervals to show the current patients' status and the output of the smart warning system implemented on the MEDIWARN Virtual Biosensor.

%%%%%%%%%%%%%%%%%%

\subsection{Requirements and Design challenges}
\label{subsec:challenges}

Communications between all the actors of the MEDIWARN system (i.e., monitors, tablets, MEDIWARN Virtual Biosensor, and monitoring station) require to design a suitable network architecture. Several requirements must be met to ensure correct operation and this raises the design challenges listed below. 

\textbf{High-bandwidth wireless communications.} The hospital units may lack of a wired network infrastructure, so it is advisable to adopt a wireless communication network. Moreover, as the monitors may need to transmit a significant amount of data within short intervals (e.g., waveforms), a network technology able to provide high network bandwidth is required.

\textbf{Mobility support.} The handheld devices have to communicate with the MEDIWARN Virtual Biosensor. As a consequence, support for node mobility has to be provided.

\textbf{Real-time operation.} The MEDIWARN Virtual Biosensor has to process data online to promptly alert the medical staff, so the network has to offer guaranteed message delivery times.

\textbf{Fault-tolerance.} The MEDIWARN system has to guarantee high availability and reliability to keep the system running even in case of faults. Hence, in the case of failure, the system has to continue operation. 
Moreover, as the vital parameters transmitted from the monitors must arrive under any circumstance (even in case of node failure, interference, etc.), suitable solutions, such as retransmissions, are needed to enable reliable communications.

%%%%%%%%%%%%%%%%%%%%

\subsection{Design choices}
\label{subsec:designChoices}

To ensure network connectivity in a typical scenario, such as the one in Fig.~\ref{fig:MEDIWARNarchitecture}, the IEEE 802.11 (WiFi) communication technology was selected. The rationale behind this design choice is that, although technologies based on the IEEE 802.15.4 standard are able to combine timeliness and reliability~\cite{7480439}, their low bandwidth and the lack of mobility support make them unsuitable to cope with the above mentioned requirements. In addition, among the available high-bandwidth wireless technologies, e.g., cellular networks, WiFi offers the advantage of being locally managed, thus entailing that the network availability does not depend on the telecom companies. Moreover, using WiFi technology any sensitive data is exchanged within the borders of the local hospital network and does not go through the Internet, thus reducing the privacy and security issues that any communication over Internet entails.
The IEEE 802.11 standard supports two operating modes, i.e., infrastructure and ad-hoc. 
Compared to WiFi working in infrastructure mode, ad-hoc WiFi provides more flexibility and higher fault tolerance, thanks to link redundancy support, which allows the transmission of the same message over multiple paths. The main disadvantage of ad-hoc WiFi is the increase of the network management complexity. In fact, the WiFi ad-hoc operation mode enables nodes to communicate when they are in the coverage range of each other. However, the wireless channel is a harsh environment, subject to interference, attenuation, and noise, so the coverage range of a node is limited. 
For these reasons, in the MEDIWARN architecture a number of intermediate nodes are placed between the monitors and the MEDIWARN Virtual Biosensor. These nodes receive the messages intended for other nodes and retransmit such messages up to their destination. This process is called message relaying and is based on a routing protocol. 
In general, network routing can be static or dynamic. In the MEDIWARN system static routing cannot be adopted for several reasons. First, the node coverage of a wireless network changes over time. Second, MEDIWARN deals with mobile nodes, i.e., the medical staff members equipped with tablets that move around the hospital unit. Moreover, when the patients to be monitored change, even the monitor location can change. 
Finally, as message delivery has to be guaranteed in any case (i.e., even in cases of intermediate node failure or corrupted transmission due to interference, noise, etc.) MEDIWARN must adopt spatial redundancy, retransmissions, and end-to-end acknowledgement. With spatial redundancy, messages are transmitted on multiple paths. End-to-end acknowledgement and retransmissions mean that upon the reception of a correct message, the destination node sends back an acknowledgement (ack). If no ack is received by the source node, the latter understands that something went wrong with the previous transmission and therefore retransmits the message. 

\section{Simulative Assessment}
\label{sec:simulativeAssessment}

This section addresses the simulative assessment of the MEDIWARN system. The simulator was developed using the OMNeT++ simulation environment and some components from the INET framework.

\subsection{Simulated scenarios}
\label{subsec:SimulatedScenarios}

\begin{figure}[htbp]
\centerline{\includegraphics[scale= 0.34]
{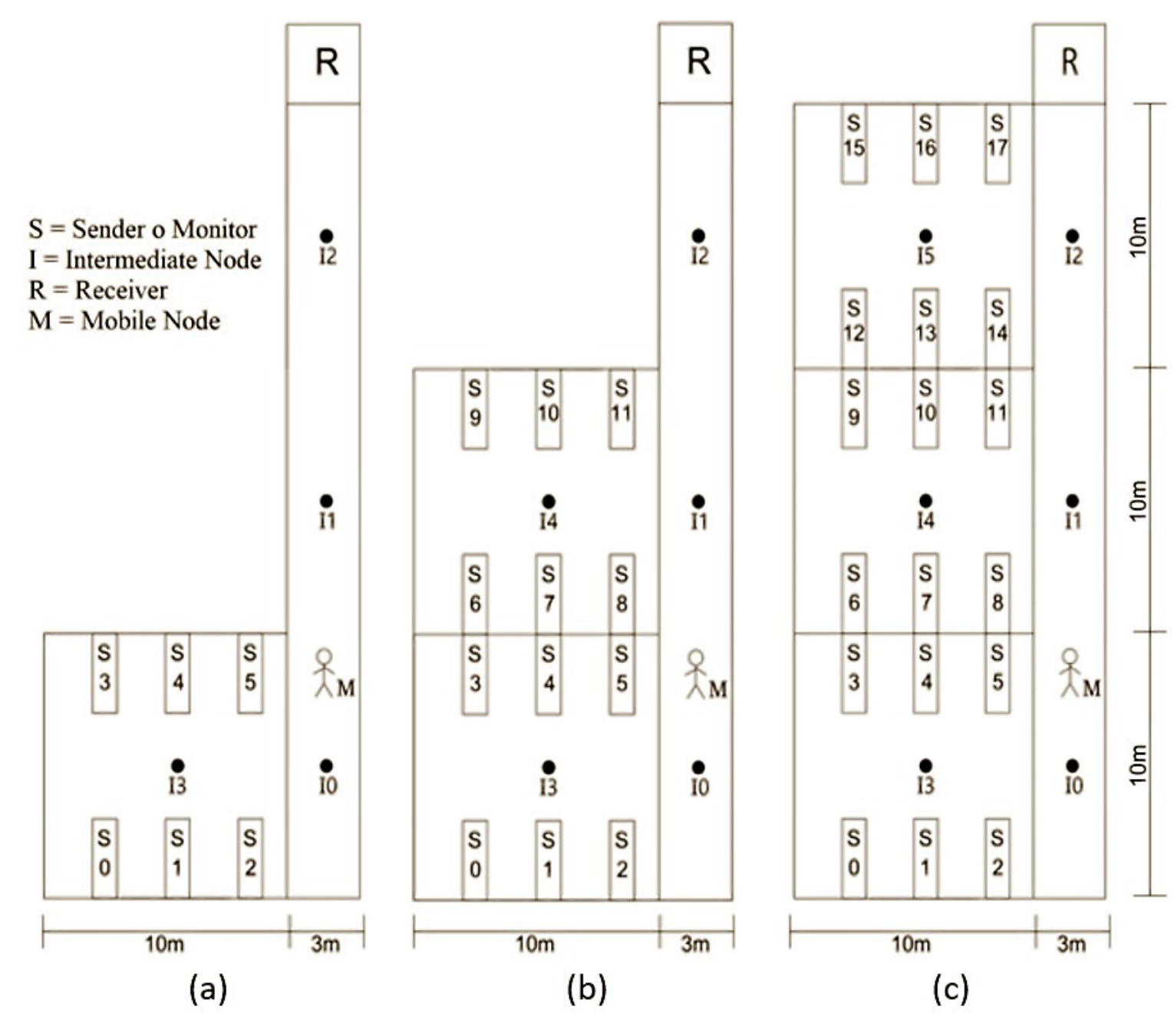}}
\caption{The simulated scenarios.}
\label{fig:scenarios}
\end{figure}

Three different scenarios, shown in Fig.~\ref{fig:scenarios}, are considered. They consist of one (\textit{scenario a}), two (\textit{scenario b}) and three (\textit{scenario c}) hospital rooms, respectively, and a 3m x 30m corridor. 
Each 10m x 10m room hosts six patients, each one with its own monitor. The monitor in our scenario represents the Sender (S), while the MEDIWARN Virtual Biosensor, which is located at the end of the corridor, is the Receiver (R). 
Some Intermediate nodes (I) are placed in the rooms and in the corridor. 
The monitors acquire the vital signals of the patients using multiple sensors and send them to the destination through multi-hop communications. Please note that both senders and intermediate nodes are stationary and perform message relaying. 
A mobile node (M) represents a doctor with his tablet, on the move in the hospital corridor. The tablet exchanges data with the MEDIWARN Virtual Biosensor through the other nodes. 
Table~\ref{tab:nodesScenario} summarizes the number and the type of nodes involved in each scenario.

%\begin{figure}[htbp]
%\centerline{\includegraphics[scale= 0.54]
%{img/scenarios1.eps}}
%\caption{The simulated scenarios.}
%\label{fig:scenarios}
%\end{figure}

\begin{table}[htbp]
\caption{Configuration of the simulated scenarios.}
\centering
\begin{tabular}{|c|c|c|c|}
\hline
 & \textbf{\textit{Scenario a}}& \textbf{\textit{Scenario b}} & \textbf{\textit{Scenario c}}\\
\hline
\textbf{Senders} & 6 & 12 & 18\\
\hline
\textbf{Mobile nodes} & 1 & 1 & 1\\
\hline
\textbf{Intermediate nodes} & 4 & 5 & 6\\
\hline
\textbf{Receivers} & 1 & 1 & 1\\
\hline
\end{tabular}
\label{tab:nodesScenario}
\end{table}

%%%%%%%%%%%%%%
%%%%%%%%%%%%%%
%%%%%%%%%%%%%%

\subsection{Simulation parameters}
\label{subsec:SimulationParameters}

The choice of the routing protocol is critical for the MEDIWARN system. The Ad-hoc On-demand Distance Vector (AODV) was identified as an interesting solution to be considered for a preliminary assessment, thanks to its features~\cite{AODV,AODVrajesh2016congestion}, i.e., it is a proactive routing algorithm that can respond very quickly to the topological change, and it has a low setup delay for connection and detection of the latest route to the destination. 
The Transmission Control Protocol (TCP) was used as Transport layer protocol. 
Each Sender generates 4 data flows, while the mobile node generates only one flow. The transmission period of each flow is 1 s. Each message is 200 byte long. The application layers of the senders are not synchronized. 
The simulation run was set to 1000 s to collect a significant amount of data.

The mobile node moves within the corridor at a speed of 2.2 m/s (i.e., a reasonable human walking speed) according to the Random Waypoint Mobility model that is available in the INET framework. 

The log-normal shadowing model with $n = 2.31$ and $\sigma = 6.42$ was adopted for the simulations. These values are realistic for indoor environments, such as an hospital~\cite{channel}. 
The relevant simulation parameters are summarized in Table~\ref{tab:simulationParameters}.

\begin{table}[htbp]
\caption{Simulation parameters.}
\centering
\begin{tabular}{|c|c|}
\hline
\textbf{Parameter} & \textbf{\textit{Value}}\\
\hline
\textbf{Speed of mobile nodes} & 2.2 m/s\\
\hline
\textbf{Physical Payload} & 200 bytes\\
\hline
\textbf{Transmission period} & 1 s\\
\hline
\textbf{Routing protocol} & AODV\\
\hline
\textbf{Propagation model} & Log-normal shadowing model\\
\hline
\textbf{Simulation time} & 1000 s\\
\hline
\end{tabular}
\label{tab:simulationParameters}
\end{table}

The used performance metric was the Round-Trip Time (RTT), here defined as the time difference between the message sending time at the sender and the reception time of the relevant ack at the same node, measured at the Application layer. 
The RTT gives an estimation of the network timings, for example, a high RTT may indicate that a message has been lost one or more times and that the message has been retransmitted one or more times, while a significant number of high RTT values may indicate a network congestion.

\subsection{Simulation results}
\label{subsec:SimulationResults}

\begin{figure}[htbp]
\centerline{\includegraphics[scale= 0.64]{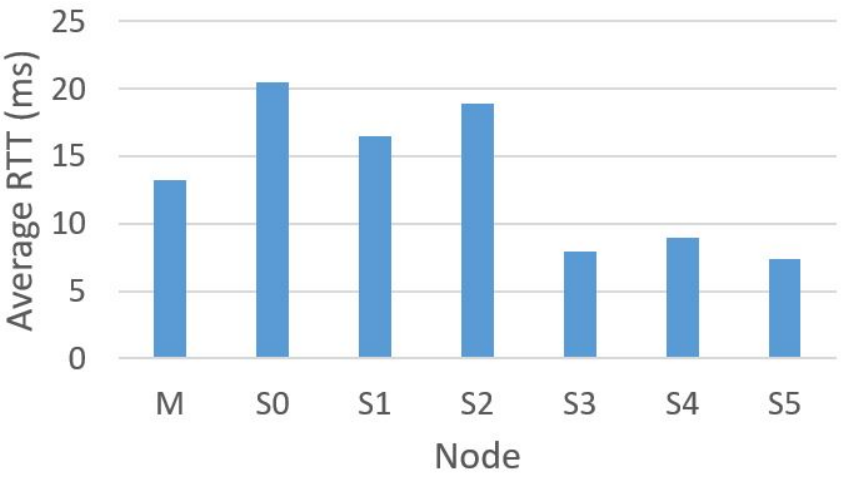}}
\caption{Scenario a - Average RTT.}
\label{fig:resultsScenarioA}
\end{figure}

\begin{figure}[htbp]
\centerline{\includegraphics[scale= 0.62]{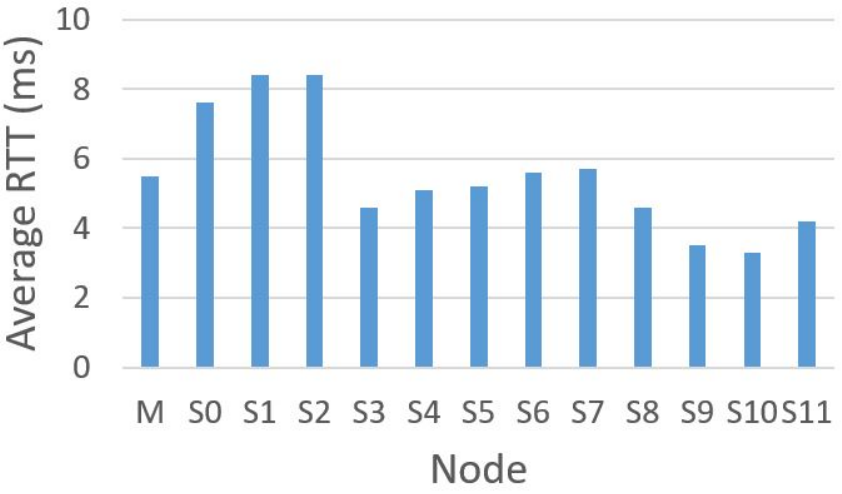}}
\caption{Scenario b - Average RTT.}
\label{fig:resultsScenarioB}
\end{figure}

\begin{figure}[htbp]
\centerline{\includegraphics[scale= 0.62]{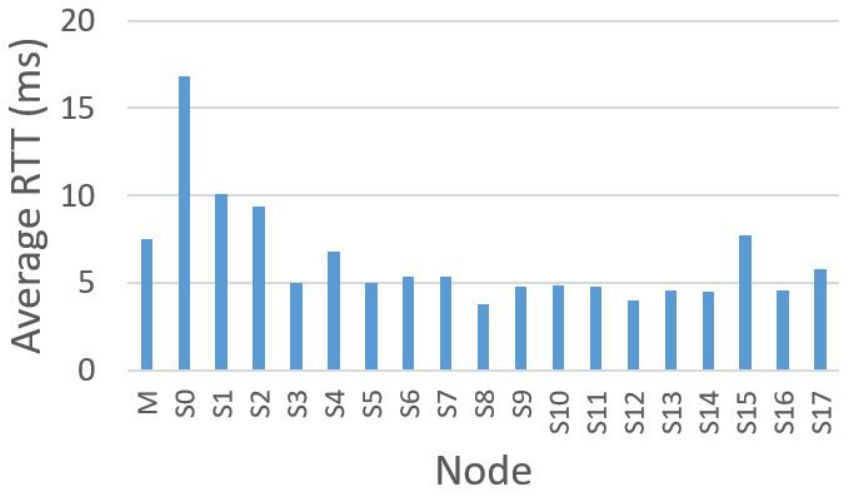}}
\caption{Scenario c - Average RTT.}
\label{fig:resultsScenarioC}
\end{figure}

Figs.~\ref{fig:resultsScenarioA},~\ref{fig:resultsScenarioB} and ~\ref{fig:resultsScenarioC} show the average RTT values obtained in the scenarios \textit{a}, \textit{b} and \textit{c}, respectively. 
The results show that the average RTT is in the order of tens of milliseconds in the assessed scenarios, therefore it stays below the transmission period. 
As expected, the RTT decreases along with the distance between the sender and the receiver. As a result, the nodes located in the rooms closer to the receiver experienced a lower RTT than the ones located in the room furthest from the receiver (i.e., the nodes S0, S1, S2, S3, S4, and S5). %As it can be seen, in the scenarios \textit{b} and \textit{c} the average RTT values are lower than the ones obtained in the scenario \textit{a} as in the latter scenario the nodes I1 and I2 are potential bottlenecks. In fact, in this scenario all the messages have to be forwarded to the receiver through the nodes I1 and I2, as they consists of the only route to reach the receiver.
As it can be seen, in the scenarios \textit{b} and \textit{c} the average RTT values are lower than the ones obtained in the scenario \textit{a}. Please note that in the latter scenario all the messages have to be forwarded to the receiver through the nodes I1 and I2, as they consists of the only route to reach the receiver. Conversely, in the other scenarios multiple routes are available to forward the messages from the senders to the receiver, as a higher number of nodes performs the message relaying. 
In the scenario \textit{c} the sender S0 measured the highest average RTT compared to the other nodes as it is the node furthest from the receiver.

%%%%%%%%%%%%%%%
%%%%%%%%%%%%%%%
%%%%%%%%%%%%%%%

In the evaluated scenarios the maximum RTT measured was about 3.55 s. This value is due the high number of retransmissions provided by the TCP when a message is not confirmed (e.g., due to interference, noise, node failure, etc.), i.e., the sender does not receive the relevant ack. 
In the MEDIWARN system the transmission period (i.e., 1 s) represents a limit value for the delivery time. However, this limit can be occasionally exceeded without compromising the proper functioning of the system. 
We note here that the data delivered to the receiver with a delay longer than the transmission period cannot be useful for real-time monitoring, as a new data sample may have been acquired at that time. 
However, these data values are important for maintaining a complete history of the patient's conditions. 
Practically, such data will not be forwarded to the monitoring station and the tablets when they poll the database server to receive new values to depict the current patients' conditions on their screen.

\section{Conclusions}
\label{sec:conclusions}

The paper described the network architecture of the MEDIWARN system and discussed some design challenges and the relevant choices to address them. We also presented a preliminary assessment of the proposed architecture using AODV as routing protocol and TCP as Transport layer protocol. 
The results obtained demonstrate that the evaluated configuration can be used for MEDIWARN but it is not an optimal solution as in some cases it does not allow the real-time monitoring and processing of sensor data due to the high message delivery time. Future work will deal with the design of a custom routing protocol for MEDIWARN. Moreover, we plan to disable the TCP protocol as we want to guarantee an upper bound on the message delivery time.  
The implementation of the MEDIWARN system on COTS devices is in progress.

%\section*{Acknowledgment}
%This research is funded by the “MEDIWARN – Virtual Biosensor for Medical Warning Precursors” European Project (CUP E69F18000000005), funded by the Programme 2014 - 2020 INTERREG V-A Italy-Malta.

\bibliographystyle{IEEEtran}
\bibliography{IEEEreferences}

\end{document}